\documentclass[twocolumn,showpacs,preprintnumbers,amsmath,amssymb,showpacs,nofootinbib, fleqn]{revtex4}

\usepackage{graphicx}
\usepackage{dcolumn}
\usepackage{color}
\usepackage{textcomp}
\usepackage{myaasmacros}

\def\bea{\begin{eqnarray}}
\def\ena{\end{eqnarray}}
\def\be{\begin{equation}}
\def\ee{\end{equation}}
\def\md{\mathrm{d}}
\def\tension{\frac{G\mu}{c^2}}
\def\anfac{\left|\sin\theta\right|}
\def\ultdm{\Delta m_0}

\begin{document}

\title{Quasar variability limits on cosmological density of cosmic strings}

\author{A.V.~Tuntsov\footnote[1]{e-mail: tyomich@sai.msu.ru}}
\affiliation{Sternberg Astronomical Institute, M.V. Lomonosov Moscow State University, 119992, Russia }
\author{M.S.~Pshirkov\footnote[2]{e-mail: pshirkov@prao.ru}}
\affiliation{Pushchino Radio Astronomy Observatory, Astro Space Center, Lebedev Physical Institute, Pushchino, 142290, Russia}

\small

\begin{abstract}
We put robust upper limits on the average cosmological density $\Omega_\mathrm{s}$ of cosmic strings based on the variability properties of a large homogeneous sample of SDSS quasars. We search for an excess of characteristic variations of quasar brightness that are associated with string lensing and use the observed distribution of this variation to constrain the density of strings. The limits obtained do not invoke any clustering of strings, apply to both open segments and closed loops of strings, usefully extend over a wide range of tensions $10^{-13}<G\mu/c^2<10^{-9}$ and reach down the level of $\Omega_\mathrm{s}=0.01$ and below. Further progress in this direction will depend on better understanding of quasar intrinsic variability rather than a mere increase in the volume of data.
\end{abstract}



\maketitle
\label{firstpage}


\section{Introduction}

Linear topological defects arise naturally during phase transitions in diverse areas of physics. Various processes in the early universe could also produce such defects, which are called cosmic strings \citep{vilenkin1994,sakellariadou2009,copeland2009}. It is often assumed that phase transitions lead to formation of strings with a characteristic tension of order the squared energy scale of the string-producing theory (in Planck units). After formation, the strings build up an intricate network, combined from open segments of the horizon scale and a multitude of loops that detach during the evolution of the network in interconnections of open strings and smoothing of their small-scale structure. The network evolves perpetually as both long segments and open loops move and oscillate at relativistic velocities.

Strings are believed to be a sub-dominant species in the matter-energy balance of the Universe. Analytical calculations and numerical simulations indicate that string networks, soon in the course of their evolution, can reach a scaling behavior where a typical distance between the strings increases in proportion to the horizon scale $d_\mathrm{h}$ \citep{allen1990, martins2004, vanchurin2005, vanchurin2006, olum2006, ringeval2007, avgoustidis2009}. This corresponds to the density of strings $\rho$ decreasing with redshift $z$ as $\rho(z)\propto d_\mathrm{h}^{-2}(z)$ although one should keep in mind that these results were obtained for radiation- and matter-dominated eras with no contribution from the vacuum energy. In the matter-dominated era this dependence coincides with that for the cold matter $\rho(z)\propto (1+z)^3$, which appears to be a natural behavior for networks dominated by non-interacting loops, whose density would decrease solely due to the universal expansion. For subsequent calculations, we will use both relations, $\rho(z)\propto d_\mathrm{h}^{-2}(z)$ and $\rho(z)\propto(1+z)^3$; the law followed by the strings in the actual universe is likely to be an interpolation between these two cases.

Despite being a natural prediction of many cosmological theories, cosmic strings have not been observed yet \cite{sazhin2003, agol2006, sazhin2007}, which raises an obvious question about the origin of this discrepancy. Attempts to answer it would benefit from an estimate of the actual density of stings in the real Universe or, in the absence of their detection, an upper limit on this parameter. However, observational estimates of this kind are surprisingly scarce \cite{morganson2009, christiansen2008, wyman2005, pogosian2004, pogosian2003}. In a  recent paper \citep{pshirkov2009}, we constrained the local (at $\sim1\,\mathrm{kpc}$ scale) density of light ($10^{-16}<G\mu/c^2<10^{-10}$) cosmic string  loops using their observational signatures in pulsar timing and precision photometric surveys. These constraints were made possible by significant enhancement in the local density of strings due to clustering of string loops expected in the Galaxy \citep{chernoff2009}. However, this enhancement is subject to theoretical uncertainties that are hard to quantify at the current level of our understanding of cosmic strings.

In the present work, we derive robust observational upper limits on the average cosmological density of cosmic strings that are independent of any clustering effects and apply to both open segments and closed loops of strings. Instead of the local enhancement, we rely on giant distances to and a large number of extragalactic objects, namely quasars, used in deriving these constraints. Our method is based on the statistical analysis of quasar variability obtained for a large sample of quasars from the SDSS catalogue \citep{vandenberk2004}. Lensing by cosmic strings that are heavy enough ($G\mu/c^2>10^{-14}-10^{-12}$, though this is somewhat quasar-model-dependent) would lead to an excess of twofold jumps in the distribution of brightness variation between two observational epochs. Hence, absence of any such features in the observed distribution allows us to infer robust upper limits on the density of strings.

The paper is organized as follows. In Section~\ref{method}, we elaborate on the idea above to see how the observed distribution of the variability in an ensemble of quasars can be used to put an upper limit on the probability of string lensing. Section~\ref{distribution_m} relates this limit to the density of strings by calculating the probability of lensing as a function of string and source parameters. Then, in Section \ref{application} we use observational data to obtain actual limits on the density of strings from the SDSS data. Finally, in Section~\ref{results} we present our results in Figure~\ref{omegafig} and conclude with a short discussion.

All numerical calculations are made for a standard flat cosmological model with a cosmological constant $\Lambda=1-\Omega$, cold matter density $\Omega=0.27$ \cite{weinberg2008} and the present-day Hubble constant $H_0=71\,\mathrm{km}\cdot\mathrm{s}^{-1}\cdot\mathrm{Mpc}^{-1}$ \citep{komatsu2009}; we assume that strings do not contribute appreciably to the energy budget.

\section{Method: Observational limits on probability of flux doubling}\label{method}

Cosmic strings produce a distinctive pattern of lensing; for a point source inside a narrow strip along the string a second positive-parity image appears in a duplicate strip on the other side of the string \cite{vilenkin1984, hogan1984, sazhin1989}. In the following, we assume that at any given moment the source is crossed by at most one string. This is a sensible approximation given that otherwise lensing by cosmic strings would be ubiquitous and would likely have been detected by now. A formal demonstration of the plausibility of this assumption relies on the fact that the strips' intersections cover a fractional area $\propto\tau^2$ in projection, where $\tau$ is the optical depth to string lensing. For cosmologically far-away sources ($z_s\sim1$), the string lensing optical depth, like that for point lenses, is of order the fraction of critical density $\Omega_s$ in the strings \citep{pshirkov2009}, which is unlikely to be greater than unity.

When an extended source being crossed by the string cannot be resolved, the observer sees a flux increase given by the flux in the part of the source that is momentarily inside the strip. As the source moves into the strip, the flux gradually rises from a flat unlensed `bottom' to some maximum value and then falls off to the same bottom as the original source disappears behind the strip while its duplicate image leaves the duplicate strip. The exact shape of the light curve depends on the brightness distribution in the source and the maximum is determined by the size of the source in relation to the strip, which can only be guessed in the case of quasars. However, for small enough sources that fit into the strip completely, the maximum increase is exactly twofold, which is $\ultdm=2.5\lg2\approx0.75^m$ in terms of stellar magnitudes. Moreover, string lensing light curves of such sources possess a characteristic extended `plateau' at this level, its width is given by the time it takes the source to traverse the strip.

The distinctive shape of the light curve readily imprints itself in the distribution density of the magnification $\mu$ of a small source lensed by a cosmic string. This function consists of a certain smooth component at $1<\mu<2$ and a pair of $\delta$-functions that correspond to the unlensed case $\mu=1$ and the maximally lensed case $\mu=2$; the latter events have a non-zero measure due to the extended nature of the corresponding `bottom' and `plateau' of the light curve.  The distribution density of the magnification allows one to calculate the density $p(\Delta m)$ of the magnitude jump $\Delta m$ in two observations due to a change in the magnification factor between the corresponding epochs. This function is even because of the symmetry between the two epochs and consists of three $\delta$-functions at $\Delta m=0$ and $\Delta m=\pm\ultdm$ on top of a smooth component $\bar{p}(\Delta m)$ for $-\ultdm<\Delta m<\ultdm$:
\begin{equation}
p(\Delta m)=\bar{p}(\Delta m)+P\delta(\Delta m\mp\ultdm)+Q\delta(\Delta m).\label{deltamdensity}
\end{equation}
For sources of unknown brightness profile, it is not possible to calculate the smooth component of this distribution; however, the amplitudes of $\delta$-functions can be calculated rather straightforwardly as shown in the next section. If lensing by cosmic strings is a rare phenomenon, which we will assume, $2P$ is essentially the optical depth (see Eq.~\ref{Ptau}) to lensing by cosmic string, which is $\tau\ll1$.

The ensemble variability studies approach the question of the variability of celestial objects by comparing the magnitudes of a large number of individual sources observed at a few (two or more) epochs and presenting various statistical measures of the individual magnitude change in the ensemble -- its mean, variance, distribution density, autocorrelation function and the like \citep{devries2006, devries2003, vandenberk2004, helfand2001, trevese1994, cimatti1993, netzer1983, macleod2008}. It is often `ergodically' assumed that these measures reflect those of individual sources to an extent given by the size and homogeneity of the observational sample and the time span of the variability survey. However, certain statistical measures of the ensemble variability have a value on their own. Of these, the distribution density $f(\Delta m)$ of the observed magnitude change will be particularly important for our study.

If the sources crossed by cosmic strings were not variable, $f(\Delta m)$ would be a direct observational estimate of the underlying density of lensing magnification $p(\Delta m)$. The quasars, on the contrary, are observed to vary at different magnitude and time scales (e.g., \cite{kozlowski2009, sesar2006, giveon1999, cimatti1993}). Nevertheless, lensing by cosmic strings might still be apparent in $f(\Delta m)$ of these objects as $\pm\ultdm$ inter-epoch changes will be overrepresented in the observed variability sample. This is the essence of our method to constrain the cosmological density of cosmic strings.

\begin{figure}
\includegraphics[width=86mm]{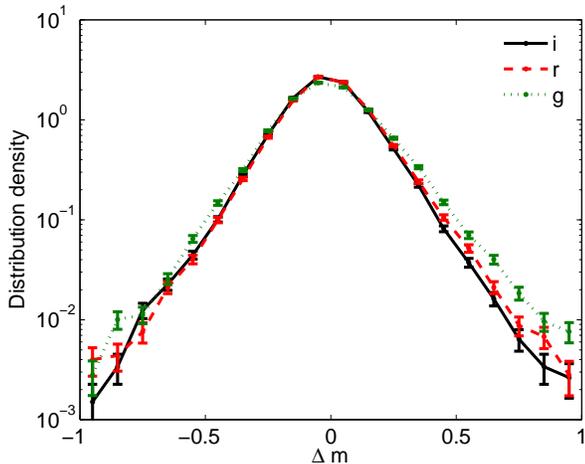}
\caption{The observed distribution densities $f(\Delta m)$ of the brightness variation $\Delta m$ between photometric and spectroscopic measurements by the Sloan Digital Sky Survey (SDSS) for $>25\,000$ quasars. The distributions in three SDSS passbands, $i$, $r$ and $g$, are derived from the results of~\cite{vandenberk2004} using a procedure described in Section~\ref{application}. The error bars shown in the figure correspond to the `Poissonian' square roots from the number of quasars in each $\Delta m$ bin.}
\label{densityfig}
\end{figure}

Mathematically, this can be expressed as follows. The string-induced variability is extrinsic to quasars and therefore the observed magnitude change distribution $f(\Delta m)$ is a convolution of those due to strings $p(\Delta m)$ and due to intrinsic processes in quasars $s(\Delta m)$\footnote{There is also a distribution of observational uncertainties but it can be absorbed into the intrinsic variability distribution; we assume that for most quasars in the sample this distribution is only weakly dependent on their actual brightness.}; for $p(\Delta m)$ given by~(\ref{deltamdensity})  the convolution equates to
\be
f(\Delta m)=Qs(\Delta m)+Ps(\Delta m\mp\ultdm))+\bar{s}(\Delta m)
\label{convolution}
\ee
with $\bar{s}(\Delta m)$ being the convolution of the intrinsic density $s(\Delta m)$ and the (unknown) smooth component of lensing density $\bar{p}(\Delta m)$. Figure~\ref{densityfig} shows an example of the observed distribution density $f(\Delta m)$ derived from the data presented in~\cite{vandenberk2004} (see Section~\ref{application} for details); it shows little evidence for any excess at $\pm\ultdm$, which will be used below to infer an upper limit on the density of cosmic strings.

Equation~(\ref{convolution}) immediately gives a handle on the parameter $P$ related to the density of cosmic strings:
\be
P=\frac{f(\Delta m)-Qs(\Delta m)-\bar{s}(\Delta m)}{s(\Delta m+\ultdm)+s(\Delta m-\ultdm)}. \label{Pgeneral}
\ee
We do {\em not} know what the intrinsic variability $s(\Delta m)$ is and therefore cannot distill the string signal from the observed $f(\Delta m)$ directly. However, if we assume that {\em all} of the variability at a certain level $\Delta m$ comes from strings, this clearly gives us an upper limit on their contribution to the variability, which can be used to infer robust constraints on the population of strings:
\be
P\le\frac{f(\Delta m)}{s(\Delta m+\ultdm)+s(\Delta m-\ultdm)}; \label{Pineq}
\ee
this inequality is valid irrespective to the assumptions on $P$ because both neglected subtrahends in the numerator of the fraction in~(\ref{Pgeneral}) are non-negative.

To deal with the denominator we assume that lensing by strings is rare; this is a sensible assumption as discussed above. In this case the amplitude $P\ll1$,  $\bar{s}(\Delta m)\ll s(\Delta m)$, $Q\approx1$, and the observed variability distribution density $f(\Delta m)$ is very close to the intrinsic one $s(\Delta m)$ -- except, possibly, at points $\Delta m=\pm\ultdm$, where a contribution $Ps(0)$  due to an excess of $\ultdm$ jumps from the lensing light curve plateau might be expected. It therefore makes sense to use~(\ref{Pineq}) at one of those points to derive an upper limit $\hat{P}$ on the parameter $P$. The denominator at these points can be approximated by the observed function $f(\Delta m)$ and one has
\be
\hat{P}\approx\frac{f(\pm\ultdm)}{f(0)+f(\pm2\ultdm)}\approx\frac{f(\pm\ultdm)}{f(0)}; \label{Pest}
\ee
the last step reflects the observational fact that $f$ measured at $\Delta m=\pm2\ultdm$ is orders of magnitude lower than at zero where it peaks (cf. Figure~\ref{densityfig}).

The constraints obtained in this way can be further refined and potentially even turned into assertive estimates for the string population properties by including additional information such as dependence of the observed distribution of magnitude change on source parameters or inter-epoch time lag. This can be accomplished by calculating the probabilities of the observed data given model parameters and using the Bayes theorem to infer the reverse. However, such an endeavor would inevitably require a model for the distribution of the intrinsic variability $s(\Delta m)$ and its dependance on source parameters, time lag or whatever else that is included in the analysis of the overall observed variability. In this study, we will use a simples approach outlined above, which is independent of authors' ignorance of the intrinsic variability of quasars though can only provide upper limits on the density of strings.

\section{Model: Probability as a function of strings population}\label{distribution_m}

The amplitude $P$ is the probability that the magnification $\Delta m$ jumps by $\ultdm$ between the two observational epochs, $t$ and $t+\Delta t$. Since $\ultdm$ is the maximum brightness increase due to string lensing, the only configuration that corresponds to this jump is that where the source is completely inside the strip in one of the observations and completely outside the strip in the other. Because of the symmetry between the two epochs we can assume that it is the first observation when the source is inside the strip and the second when it is outside thereby replacing $\Delta t$ with is absolute value:
\begin{equation}
2P=\mathcal{P}\left[\Delta m(t)=\ultdm~\mathrm{and}~\Delta m(t+|\Delta t|)=0\right]. \label{def2P}
\end{equation}
To estimate this value we first introduce the angular width $\Delta$ of the string lensing strip. According to \cite{vilenkin1984, hogan1984}, it depends on the tension $\mu$ of the string  and its local inclination $\theta$ to the line of sight :
\begin{equation}
\Delta=8\pi\anfac\tension\frac{D_\mathrm{ls}}{D_\mathrm{os}} \label{stripwidth},
\end{equation}
where $D_{os}$ and $D_{ls}$ are the (angular diameter) distances, respectively, from the observer and from the string to the source (along the line of sight); we use the average value of $\langle\anfac\rangle=\pi/4$. It seems sufficient for our study to assume that the string segment responsible for lensing is long and straight compared to the angular size of the source; for a comprehensive study of lensing by general configurations of strings see \cite{uzan2001, bernardeau2001}.

Now let $x$ be the initial epoch position of the projection of the source center onto the lens plane with respect to the strip median line (measured towards the outer edge of the strip such that the string itself is at $x_s=-\Delta/2$). The position of the source in the second epoch is then $x+\beta_\perp c|\Delta t|/(1+z_\mathrm{l}) D_\mathrm{ol}$, where $\beta_\perp c$ is the orthogonal (to the string) component of transverse (to the line of sight) velocity of the string w.r.t. the source; factor $(1+z_\mathrm{l})^{-1}$ corresponds to the dilation of observed time lag  $\Delta t$ from the lens plane at redshift $z_\mathrm{l}$. Cosmic strings are expected to move relativistically, $\beta\sim\mathcal{O}(1)$ \cite{vilenkin1994}; following \cite{kuijken2008, pshirkov2009}, we  use $\beta_\perp=0.3$ in subsequent calculations\footnote{We assume that $\beta_\perp>0$, i.e. the source is moving away from the string; this is not restrictive due to time symmetry mentioned.}.

Since we assume that the source is lensed by at most one string, conditions in the argument of~(\ref{def2P}) require that the center of the source is inside the strip by a margin of at least the source size $r_\perp=R_\perp/D_\mathrm{os}$ ($R_\perp$ is its linear size) on the first observation and outside it by the same margin on the second epoch:
\begin{equation}
\left\{
\begin{array}{l}
|x|\le\Delta/2-r_\perp\\
x+\beta_\perp c|\Delta t|/(1+z_\mathrm{l})D_\mathrm{ol}\ge\Delta/2+r_\perp
\end{array}
\right.. \label{xconditions}
\end{equation}
Taken together, they restrict $x$ to a narrow strip of width $\xi$, which is the lowest of $\Delta -2r_\perp$ and $\beta_\perp c|\Delta t|/(1+z_\mathrm{l})D_\mathrm{ol}-2r_\perp$ as long as this lowest is positive, and zero otherwise:
\begin{equation}
\xi=\mathrm{max}\left\{0, \mathrm{min}\left[\Delta , \frac{\beta_\perp c|\Delta t|}{(1+z_\mathrm{l})D_\mathrm{ol}}\right]-2r_\perp\right\}. \label{xidef}
\end{equation}
The probability that a randomly placed source will lie within the strip of this width parallel to a string in an infinitesimally thin slice of string network with local number density $\rho/\mu$ is
\begin{equation}
\md\tau=\frac\rho\mu\xi D_\mathrm{ol}\md \bar{D}_\mathrm{ol} =\Omega_\mathrm{s}\frac{3H_0^2}{8\pi G\mu}\omega(z_\mathrm{l})\xi D_\mathrm{ol}\md \bar{D}_\mathrm{ol}, \label{dtau}
\end{equation}
where $\bar{D}$ is the proper distance along the line of sight parameterized by the slice redshift $z_\mathrm{l}$.

In the formula above we also introduced the current cosmological density of strings $\Omega_\mathrm{s}$ and its dependence on redshift $\omega(z_\mathrm{l})$ such that the proper density $\rho(z_\mathrm{l})=\omega(z_\mathrm{l})\Omega_\mathrm{s} 3H_0^2/8\pi G$. We use two models for $\omega(z_\mathrm{l})$ -- that corresponding to scaling solutions
\be
\omega(z)=\left[\frac{d_\mathrm{h}(0)}{d_\mathrm{h}(z)}\right]^{2}, \label{rhoscaling}
\ee
and pressureless dust:
\be
\omega(z)=(1+z)^3 \label{rhodust}.
\ee
As discussed in the Introduction, there is currently no consensus on the relative contribution of closed and open strings to the energy budget of the Universe, but whatever the contributions are, the final result for the density constraints can be obtained by interpolating those derived from the application of~(\ref{rhoscaling}) and~(\ref{rhodust}).

With equations~(\ref{xidef}, \ref{dtau}) we can now write down the probability $2P$ of a twofold magnification jump in any of the infinitesimal slices. It is then given by the integrated optical depth $\tau$ along the line of sight to the source
\be
\frac\tau{\Omega_\mathrm{s}}=\frac{3H_0^2}{8\pi G\mu}\int\limits_0^{z_\mathrm{s}}\md \bar{D}(z_\mathrm{l})\, \omega(z_\mathrm{l}) D(z_\mathrm{l})\xi\left(z_\mathrm{l}, z_\mathrm{s}, \mu, R_\perp, |\Delta t|\right) \label{taudef}
\ee
according to
\begin{equation}
\label{Ptau}
P=\frac12\left(1-e^{-\tau}\right)=\frac\tau2+\mathcal{O}(\tau^2);
\end{equation}
this probability is placed symmetrically in $\Delta m=\pm\ultdm$, hence the factor $1/2$ in front of the  brackets.

\section{Application: Observational limits on string density}\label{application}

In order to put an upper limit $\hat{\Omega}_\mathrm{s}$ on the density of cosmic strings one now can simply equate the observational upper limit $\hat{P}$ on the probability of lensing to its model estimate $P$ given by~(\ref{Ptau}) in the limit $\tau\ll1$:
\begin{equation}
\hat{\Omega}_\mathrm{s}=\frac{2\hat{P}}{\tau/\Omega_\mathrm{s}}. \label{omegaest}
\end{equation}
The numerator of the fraction above can be estimated using~(\ref{Pest}) from the distribution density $f(\Delta m)$ of quasar brightness variations, which can be calculated directly from the observational data. In this regard, SDSS quasar survey  \cite{york2000} provides an invaluable observational sample, where the brightness of tens of thousands of quasars is homogeneously measured in a number of optical passbands and could be compared against an equally homogeneous sample of brightness estimates derived from the quasar spectra, which are obtained month and years after the photometric observations.

Such an analysis has indeed been done for $N=25710$ SDSS quasars by Vanden Berk et~al.~\cite{vandenberk2004} and we use their data on quasar variability in SDSS passbands $g$, $r$ and $i$ to derive constraints on the string population. The authors of the cited study do not explicitly quote estimates on the probability distribution density $f(\Delta m)$ and we do not possess sufficient resource to re-reduce the publicly available SDSS data to derive $f(\Delta m)$ independently but it can be readily obtained from Figure~3 in the PDF version of~\cite{vandenberk2004}. To do so, we manually counted the data points corresponding to individual quasar measurements in the scatter plots of the figure in $0.1^m$-wide bins for $|\Delta m|\ge 0.4^m$ or summed the heights of the respective histograms for $|\Delta m|<0.4^m$ \footnote{The quality of the figure does not allow us to use a consistent counting approach in the entire domain of $\Delta m$ -- inner regions ($|\Delta m|<0.4^m$) of scatter plots suffer from considerable confusion of data points while the linear scale of the histograms makes them hardly readable for $|\Delta m|>0.6^m$. However, where this comparison is possible, at bins centered at $\pm0.45^m$ and $\pm0.55^m$, the numbers  agree to within a few percent, which is acceptable given somewhat low-tech approach employed in the absence of published digital data.}, and then divided them by the bin width and the total number of quasars in the ensemble.

The distribution densities of quasar variability in three passbands obtained in this way are plotted in Figure~\ref{densityfig} while Table~\ref{pesttable} presents corresponding values $f(\Delta m)$ at points of our interest and an estimate for $\hat{P}$ in each passband. Since lensing is achromatic (as long as the string is heavy enough, such that $\Delta\ge 2r_\perp$ for all passbands) and intrinsic variability is not, we are free to choose the lowest $\hat{P}$ to use in~(\ref{omegaest}); this is $\hat{P}=3.2\cdot10^{-3}$, which corresponds to passband $r$.

\begin{table}
\caption{Observational estimates for the distribution density of quasar brightness variation at $\Delta m=0,\pm\ultdm$ in three SDSS passbands and corresponding estimates for $\hat{P}$. For consistency, we calculate the central value of the distribution density $f(0^m)=[f(-0.05^m)+f(0.05^m)]/2$; the value of $\hat{P}$ quoted in the table is the average between values corresponding to $\Delta m=-\ultdm$ and $\Delta m=\ultdm$ according to~(\ref{Pest}). }
\label{pesttable}
\begin{tabular}{ccccccccc}
Passband & $f(0^m)$ & & $f(-0.75^m)$ &  & $f(0.75^m)$ & & $\hat{P}$\\
\hline\\
$g$ & $2.2$ &  & $1.1\cdot10^{-2}$ & & $1.8\cdot10^{-2}$ & & $6.6\cdot10^{-3}$\\
$r$ & $2.5$ & & $7.6\cdot10^{-3}$ & & $8.8\cdot10^{-3}$ & & $3.2\cdot10^{-3}$\\
$i$ & $2.5$ & & $1.2\cdot10^{-2}$ &  & $6.4\cdot10^{-3}$ & & $3.7\cdot10^{-3}$\\
\hline
\end{tabular}
\end{table}

The denominator of the fraction in~(\ref{omegaest}) is given by the right-hand side of~(\ref{taudef}), which depends on the source redshift, and we therefore need to take an average with respect to the distribution of the observed quasars. The quasar sample of~\cite{vandenberk2004} includes most of the quasars in SDSS Data Release 1 Quasar Catalogue \cite{schneider2003} and a substantial fraction of quasars observed by SDSS that were not included in SDSS DR1. The properties of individual quasars in the entire sample used in that study do not appear to have ever been detailed in a publication and therefore we use SDSS DR3 QSO catalogue \cite{schneider2005} as a proxy to the statistical properties of the true sample. We have verified numerically, that our results do not change significantly if we use either DR1 or DR5 \cite{schneider2007} catalogues in averaging $\tau/\Omega_\mathrm{s}$ (numbers do get higher when using more recent, deeper versions of the catalogue but only by 2--3 per cent).

Another source of uncertainty is the physical size $2R_\perp$ that produces most of the observed  flux. The size of the quasar affects our results significantly by limiting the sensitivity of our estimate as a function of string tension $\mu$. Estimates on these quantities vary appreciably in the literature, depending on the method used; reverberation mapping seems to favor sizes in the range $R\sim(10^{16} - 10^{17})\,\mathrm{cm}$ \citep{bentz2009, melnikov2008, peterson2004, wandel1999, kaspi2000} while microlensing techniques  give somewhat smaller values of $R\sim(10^{15} - 10^{16})\,\mathrm{cm}$ \citep{kochanek2004, floyd2008, vakulik2007, wayth2005, wyithe2002}. Both methods are not model-independent and the estimates are expected to correlate with individual properties of QSOs such as the luminosity or the mass of the central black hole. In the apparent absence of a better option, we perform our calculations using three representative values of $2R_\perp\in\{10^{15}, 10^{16}, 10^{17}\}\,\mathrm{cm}$ treating $10^{16}\,\mathrm{cm}$ as a fiducial estimate.

Finally, the value of the time lag $\Delta t$ between two observational epochs in the observer frame cannot be read from the results of \cite{vandenberk2004} directly, which also introduces some uncertainty. The value of $\Delta t$ affects our results directly via~(\ref{xidef}) and therefore need to be fixed to perform calculations. From the visual inspection of Figure~4 of~\cite{vandenberk2004} it is clear that typical time lag in the source frame $\Delta t/(1+z_\mathrm{s})\sim (100 - 200)\,\mathrm{days}$. The average redshift of SDSS quasars is $z_\mathrm{s}\approx 1.5$. We therefore take $\Delta t=150\,\mathrm{days}\times2.5\approx 3.2\cdot10^7\,\mathrm{s}$.

\section{Results and Discussion}\label{results}

\begin{figure}
\center
\includegraphics[width=86mm]{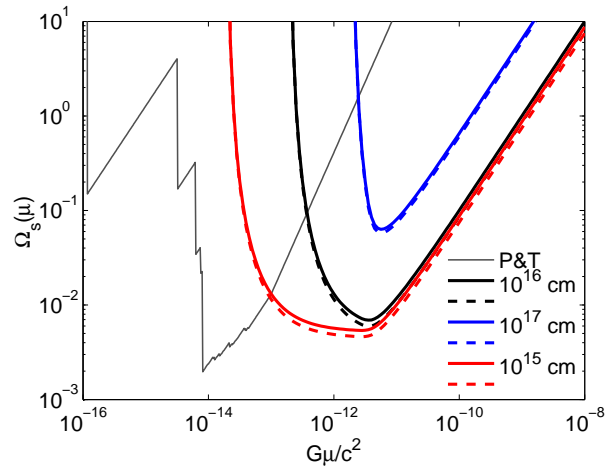}
\caption{Upper limit on the average present-day cosmological density of light cosmic strings as a function of its tension $\mu$ based on the quasar variability distribution shown in Figure~\ref{densityfig}. Three pairs of curves are shown for the assumed source size $2R_\perp$ of $10^{16}\,\mathrm{cm}$ (black, middle), $10^{17}\,\mathrm{cm}$ (blue, top) and $10^{15}\,\mathrm{cm}$ (red, bottom). Solid lines correspond to `scaling' evolution of string density with redshift according to~(\ref{rhoscaling}), dashed ones assume a `dust-like' law~(\ref{rhodust}); they do not differ much. A thin grey line shows the constraints obtained in~\cite{pshirkov2009} from the local effects of strings.}
\label{omegafig}
\end{figure}

Figure~\ref{omegafig} presents the upper limits on the average cosmological density of strings set by the statistics of the observed variability of more than 25 thousand SDSS quasars as a function of string tension $G\mu/c^2$. Depending on the assumed size of the source, which turns to be the major parameter of the method developed in this paper, the region where the strings density is usefully constrained extends for up to five orders of magnitude. At the same time, our constraints are only weakly dependent on the assumed behavior of string density with redshift, mostly because prescriptions~(\ref{rhoscaling}) and~(\ref{rhodust}) started to diverge from each other relatively recently, when the vacuum density began to dominate in the Universe.

The most stringent constraints are obtained at $G\mu/c^2\sim (10^{-13} -10^{-11})$ where the upper limits reach below the level of $\Omega_\mathrm{s}=0.01$. We note that these are rather weak limits for open topological cosmic strings because their cosmological density is believed to be of order $\Omega_\mathrm{s}\sim (10-100) G\mu/c^2$ based on the reconnection argument leading to scaling solutions (e.g., \cite{sakellariadou2009}). However, this argument might not be applicable to fundamental strings, for which the reconnection probability is highly model dependent and can be significantly lower than unity \cite{jackson2005, sakellariadou2005}; neither it is clear how it applies to cosmic string loops, which are essentially a product of the mechanism that ensures that the density of open string has a scaling behavior. Our observational upper limits apply to all kinds of strings, whether topological or fundamental, open or otherwise, and therefore provide strong constraints in the case of string loops and fundamental strings, independent of theoretical uncertainty. Moreover, our results might potentially be useful for constraining the unknown reconnection probability of fundamental strings.

The upper limits obtained here rely on an extensive dataset of SDSS quasars and are robust in the sense that they do not depend on any local enhancement of the density expected for string loops in the Galaxy \cite{chernoff2009} or a particular model of quasar emission, as long as the source remains sufficiently small compared to the lensing strip width. This method is also rather insensitive to the photometric accuracy of observations because $0.75^m$-wide jumps in brightness are fairly obvious by any standard. And if follow-up observations could be performed when a survey telescope sees a sudden twofold increase in the brightness of a quasar, it might even be possible to confirm or rule out string nature of this increase to a high level of confidence.

However, all this robustness also means that upper limits obtained in this paper could not be improved much by simply increasing the number of data points in the sample, which is expected to grow by orders of magnitude with the launch of next generation photometric surveys, such as LSST \cite{lsst2009}. Rather, it is the study of quasar intrinsic variability that can bring most benefit for this cause. The variation distribution density presented in Figure~\ref{densityfig} does not appear to show any notable features around $\Delta m=\pm0.75^m$ which would make these points stand our from the smooth decrease of $f(\Delta m)$ with $|\Delta m|$. Therefore, the upper limits derived in this paper are most likely a significant over-estimation of the true density of strings because what is conservatively interpreted as a string `signal' here is most likely the quasar variability `noise'. It will not simply walk away if one takes more quasars. To make further progress here, one needs to learn to separate two contributions -- e.g., using the varying dependence of the two effects on various parameters, such as the source redshift, color, luminosity and so on. We can easily calculate how the statistics of string lensing effect should depend on these parameters but at present are less certain when it comes to quasar intrinsic variability.

Nevertheless, the science of quasar variability is advancing fast and our understanding of it might soon be sufficient for distilling the string effect from the observational data. This approach can also be followed in the analysis of future datasets from the GAIA mission \cite{perryman2001} on the variability of stars, which are less distant but photometrically more stable and much more numerous than SDSS quasars. Moreover, at present we do not know the density of cosmic strings in the Universe and cannot predict just when the twin peaks of strings signal at $\Delta m=\pm\ultdm$ will begin to show up in the growing datasets of ensemble variability studies.


\section*{Acknowledgements}

This paper is based on the data collected in the course of the Sloan Digital Sky Survey project and made suitable for an application of our method by the authors of~\cite{vandenberk2004}. Funding for the Sloan Digital Sky Survey (SDSS) has been provided by the Alfred P. Sloan Foundation, the Participating Institutions, the National Aeronautics and Space Administration, the National Science Foundation, the U.S. Department of Energy, the Japanese Monbukagakusho, and the Max Planck Society. The SDSS Web site is http://www.sdss.org/.

The SDSS is managed by the Astrophysical Research Consortium (ARC) for the Participating Institutions. The Participating Institutions are The University of Chicago, Fermilab, the Institute for Advanced Study, the Japan Participation Group, The Johns Hopkins University, Los Alamos National Laboratory, the Max-Planck-Institute for Astronomy (MPIA), the Max-Planck-Institute for Astrophysics (MPA), New Mexico State University, University of Pittsburgh, Princeton University, the United States Naval Observatory, and the University of Washington.

Use of NASA's Astrophysics Data System is also gratefully acknowledged. We thank the referee for an important suggestion.


\bibliography{qsostrings}{}

\begin{thebibliography}{57}
\expandafter\ifx\csname natexlab\endcsname\relax\def\natexlab#1{#1}\fi
\expandafter\ifx\csname bibnamefont\endcsname\relax
  \def\bibnamefont#1{#1}\fi
\expandafter\ifx\csname bibfnamefont\endcsname\relax
  \def\bibfnamefont#1{#1}\fi
\expandafter\ifx\csname citenamefont\endcsname\relax
  \def\citenamefont#1{#1}\fi
\expandafter\ifx\csname url\endcsname\relax
  \def\url#1{\texttt{#1}}\fi
\expandafter\ifx\csname urlprefix\endcsname\relax\def\urlprefix{URL }\fi
\providecommand{\bibinfo}[2]{#2}
\providecommand{\eprint}[2][]{\url{#2}}

\bibitem[{\citenamefont{{Vilenkin} and {Shellard}}(1994)}]{vilenkin1994}
\bibinfo{author}{\bibfnamefont{A.}~\bibnamefont{{Vilenkin}}} \bibnamefont{and}
  \bibinfo{author}{\bibfnamefont{E.~P.~S.} \bibnamefont{{Shellard}}},
  \emph{\bibinfo{title}{{Cosmic strings and other topological defects}}}
  (\bibinfo{year}{1994}).

\bibitem[{\citenamefont{{Sakellariadou}}(2009)}]{sakellariadou2009}
\bibinfo{author}{\bibfnamefont{M.}~\bibnamefont{{Sakellariadou}}},
  \bibinfo{journal}{Nuclear Physics B Proceedings Supplements}
  \textbf{\bibinfo{volume}{192}}, \bibinfo{pages}{68} (\bibinfo{year}{2009}),
  \eprint{0902.0569}.

\bibitem[{\citenamefont{{Copeland} and {Kibble}}(2009)}]{copeland2009}
\bibinfo{author}{\bibfnamefont{E.~J.} \bibnamefont{{Copeland}}}
  \bibnamefont{and} \bibinfo{author}{\bibfnamefont{T.~W.~B.}
  \bibnamefont{{Kibble}}}, \bibinfo{journal}{ArXiv e-prints}
  (\bibinfo{year}{2009}), \eprint{0911.1345}.

\bibitem[{\citenamefont{{Allen} and {Shellard}}(1990)}]{allen1990}
\bibinfo{author}{\bibfnamefont{B.}~\bibnamefont{{Allen}}} \bibnamefont{and}
  \bibinfo{author}{\bibfnamefont{E.~P.~S.} \bibnamefont{{Shellard}}},
  \bibinfo{journal}{Physical Review Letters} \textbf{\bibinfo{volume}{64}},
  \bibinfo{pages}{119} (\bibinfo{year}{1990}).

\bibitem[{\citenamefont{{Martins}}(2004)}]{martins2004}
\bibinfo{author}{\bibfnamefont{C.~J.} \bibnamefont{{Martins}}},
  \bibinfo{journal}{\prd} \textbf{\bibinfo{volume}{70}},
  \bibinfo{pages}{107302} (\bibinfo{year}{2004}),
  \eprint{arXiv:hep-ph/0410326}.

\bibitem[{\citenamefont{{Vanchurin} et~al.}(2005)\citenamefont{{Vanchurin},
  {Olum}, and {Vilenkin}}}]{vanchurin2005}
\bibinfo{author}{\bibfnamefont{V.}~\bibnamefont{{Vanchurin}}},
  \bibinfo{author}{\bibfnamefont{K.}~\bibnamefont{{Olum}}}, \bibnamefont{and}
  \bibinfo{author}{\bibfnamefont{A.}~\bibnamefont{{Vilenkin}}},
  \bibinfo{journal}{\prd} \textbf{\bibinfo{volume}{72}},
  \bibinfo{pages}{063514} (\bibinfo{year}{2005}), \eprint{arXiv:gr-qc/0501040}.

\bibitem[{\citenamefont{{Vanchurin} et~al.}(2006)\citenamefont{{Vanchurin},
  {Olum}, and {Vilenkin}}}]{vanchurin2006}
\bibinfo{author}{\bibfnamefont{V.}~\bibnamefont{{Vanchurin}}},
  \bibinfo{author}{\bibfnamefont{K.~D.} \bibnamefont{{Olum}}},
  \bibnamefont{and}
  \bibinfo{author}{\bibfnamefont{A.}~\bibnamefont{{Vilenkin}}},
  \bibinfo{journal}{\prd} \textbf{\bibinfo{volume}{74}},
  \bibinfo{pages}{063527} (\bibinfo{year}{2006}), \eprint{arXiv:gr-qc/0511159}.

\bibitem[{\citenamefont{{Olum} and {Vilenkin}}(2006)}]{olum2006}
\bibinfo{author}{\bibfnamefont{K.~D.} \bibnamefont{{Olum}}} \bibnamefont{and}
  \bibinfo{author}{\bibfnamefont{A.}~\bibnamefont{{Vilenkin}}},
  \bibinfo{journal}{\prd} \textbf{\bibinfo{volume}{74}},
  \bibinfo{pages}{063516} (\bibinfo{year}{2006}),
  \eprint{arXiv:astro-ph/0605465}.

\bibitem[{\citenamefont{{Ringeval} et~al.}(2007)\citenamefont{{Ringeval},
  {Sakellariadou}, and {Bouchet}}}]{ringeval2007}
\bibinfo{author}{\bibfnamefont{C.}~\bibnamefont{{Ringeval}}},
  \bibinfo{author}{\bibfnamefont{M.}~\bibnamefont{{Sakellariadou}}},
  \bibnamefont{and} \bibinfo{author}{\bibfnamefont{F.~R.}
  \bibnamefont{{Bouchet}}}, \bibinfo{journal}{Journal of Cosmology and
  Astro-Particle Physics} \textbf{\bibinfo{volume}{2}}, \bibinfo{pages}{23}
  (\bibinfo{year}{2007}), \eprint{arXiv:astro-ph/0511646}.

\bibitem[{\citenamefont{{Avgoustidis} and {Copeland}}(2009)}]{avgoustidis2009}
\bibinfo{author}{\bibfnamefont{A.}~\bibnamefont{{Avgoustidis}}}
  \bibnamefont{and} \bibinfo{author}{\bibfnamefont{E.~J.}
  \bibnamefont{{Copeland}}}, \bibinfo{journal}{ArXiv e-prints}
  (\bibinfo{year}{2009}), \eprint{0912.4004}.

\bibitem[{\citenamefont{{Sazhin} et~al.}(2003)\citenamefont{{Sazhin}, {Longo},
  {Capaccioli}, {Alcal{\'a}}, {Silvotti}, {Covone}, {Khovanskaya}, {Pavlov},
  {Pannella}, {Radovich} et~al.}}]{sazhin2003}
\bibinfo{author}{\bibfnamefont{M.}~\bibnamefont{{Sazhin}}},
  \bibinfo{author}{\bibfnamefont{G.}~\bibnamefont{{Longo}}},
  \bibinfo{author}{\bibfnamefont{M.}~\bibnamefont{{Capaccioli}}},
  \bibinfo{author}{\bibfnamefont{J.~M.} \bibnamefont{{Alcal{\'a}}}},
  \bibinfo{author}{\bibfnamefont{R.}~\bibnamefont{{Silvotti}}},
  \bibinfo{author}{\bibfnamefont{G.}~\bibnamefont{{Covone}}},
  \bibinfo{author}{\bibfnamefont{O.}~\bibnamefont{{Khovanskaya}}},
  \bibinfo{author}{\bibfnamefont{M.}~\bibnamefont{{Pavlov}}},
  \bibinfo{author}{\bibfnamefont{M.}~\bibnamefont{{Pannella}}},
  \bibinfo{author}{\bibfnamefont{M.}~\bibnamefont{{Radovich}}},
  \bibnamefont{et~al.}, \bibinfo{journal}{\mnras}
  \textbf{\bibinfo{volume}{343}}, \bibinfo{pages}{353} (\bibinfo{year}{2003}),
  \eprint{arXiv:astro-ph/0302547}.

\bibitem[{\citenamefont{{Agol} et~al.}(2006)\citenamefont{{Agol}, {Hogan}, and
  {Plotkin}}}]{agol2006}
\bibinfo{author}{\bibfnamefont{E.}~\bibnamefont{{Agol}}},
  \bibinfo{author}{\bibfnamefont{C.~J.} \bibnamefont{{Hogan}}},
  \bibnamefont{and} \bibinfo{author}{\bibfnamefont{R.~M.}
  \bibnamefont{{Plotkin}}}, \bibinfo{journal}{\prd}
  \textbf{\bibinfo{volume}{73}}, \bibinfo{pages}{087302}
  (\bibinfo{year}{2006}), \eprint{arXiv:astro-ph/0603838}.

\bibitem[{\citenamefont{{Sazhin} et~al.}(2007)\citenamefont{{Sazhin},
  {Khovanskaya}, {Capaccioli}, {Longo}, {Paolillo}, {Covone}, {Grogin}, and
  {Schreier}}}]{sazhin2007}
\bibinfo{author}{\bibfnamefont{M.~V.} \bibnamefont{{Sazhin}}},
  \bibinfo{author}{\bibfnamefont{O.~S.} \bibnamefont{{Khovanskaya}}},
  \bibinfo{author}{\bibfnamefont{M.}~\bibnamefont{{Capaccioli}}},
  \bibinfo{author}{\bibfnamefont{G.}~\bibnamefont{{Longo}}},
  \bibinfo{author}{\bibfnamefont{M.}~\bibnamefont{{Paolillo}}},
  \bibinfo{author}{\bibfnamefont{G.}~\bibnamefont{{Covone}}},
  \bibinfo{author}{\bibfnamefont{N.~A.} \bibnamefont{{Grogin}}},
  \bibnamefont{and} \bibinfo{author}{\bibfnamefont{E.~J.}
  \bibnamefont{{Schreier}}}, \bibinfo{journal}{\mnras}
  \textbf{\bibinfo{volume}{376}}, \bibinfo{pages}{1731} (\bibinfo{year}{2007}),
  \eprint{arXiv:astro-ph/0611744}.

\bibitem[{\citenamefont{{Morganson} et~al.}(2009)\citenamefont{{Morganson},
  {Marshall}, {Treu}, {Schrabback}, and {Blandford}}}]{morganson2009}
\bibinfo{author}{\bibfnamefont{E.}~\bibnamefont{{Morganson}}},
  \bibinfo{author}{\bibfnamefont{P.}~\bibnamefont{{Marshall}}},
  \bibinfo{author}{\bibfnamefont{T.}~\bibnamefont{{Treu}}},
  \bibinfo{author}{\bibfnamefont{T.}~\bibnamefont{{Schrabback}}},
  \bibnamefont{and} \bibinfo{author}{\bibfnamefont{R.~D.}
  \bibnamefont{{Blandford}}}, \bibinfo{journal}{ArXiv e-prints}
  (\bibinfo{year}{2009}), \eprint{0908.0602}.

\bibitem[{\citenamefont{{Christiansen}
  et~al.}(2008)\citenamefont{{Christiansen}, {Albin}, {James}, {Goldman},
  {Maruyama}, and {Smoot}}}]{christiansen2008}
\bibinfo{author}{\bibfnamefont{J.~L.} \bibnamefont{{Christiansen}}},
  \bibinfo{author}{\bibfnamefont{E.}~\bibnamefont{{Albin}}},
  \bibinfo{author}{\bibfnamefont{K.~A.} \bibnamefont{{James}}},
  \bibinfo{author}{\bibfnamefont{J.}~\bibnamefont{{Goldman}}},
  \bibinfo{author}{\bibfnamefont{D.}~\bibnamefont{{Maruyama}}},
  \bibnamefont{and} \bibinfo{author}{\bibfnamefont{G.~F.}
  \bibnamefont{{Smoot}}}, \bibinfo{journal}{\prd}
  \textbf{\bibinfo{volume}{77}}, \bibinfo{pages}{123509}
  (\bibinfo{year}{2008}), \eprint{0803.0027}.

\bibitem[{\citenamefont{{Wyman} et~al.}(2005)\citenamefont{{Wyman}, {Pogosian},
  and {Wasserman}}}]{wyman2005}
\bibinfo{author}{\bibfnamefont{M.}~\bibnamefont{{Wyman}}},
  \bibinfo{author}{\bibfnamefont{L.}~\bibnamefont{{Pogosian}}},
  \bibnamefont{and}
  \bibinfo{author}{\bibfnamefont{I.}~\bibnamefont{{Wasserman}}},
  \bibinfo{journal}{\prd} \textbf{\bibinfo{volume}{72}},
  \bibinfo{pages}{023513} (\bibinfo{year}{2005}),
  \eprint{arXiv:astro-ph/0503364}.

\bibitem[{\citenamefont{{Pogosian} et~al.}(2004)\citenamefont{{Pogosian},
  {Wyman}, and {Wasserman}}}]{pogosian2004}
\bibinfo{author}{\bibfnamefont{L.}~\bibnamefont{{Pogosian}}},
  \bibinfo{author}{\bibfnamefont{M.}~\bibnamefont{{Wyman}}}, \bibnamefont{and}
  \bibinfo{author}{\bibfnamefont{I.}~\bibnamefont{{Wasserman}}},
  \bibinfo{journal}{Journal of Cosmology and Astro-Particle Physics}
  \textbf{\bibinfo{volume}{9}}, \bibinfo{pages}{8} (\bibinfo{year}{2004}),
  \eprint{arXiv:astro-ph/0403268}.

\bibitem[{\citenamefont{{Pogosian} et~al.}(2003)\citenamefont{{Pogosian},
  {Tye}, {Wasserman}, and {Wyman}}}]{pogosian2003}
\bibinfo{author}{\bibfnamefont{L.}~\bibnamefont{{Pogosian}}},
  \bibinfo{author}{\bibfnamefont{S.}~\bibnamefont{{Tye}}},
  \bibinfo{author}{\bibfnamefont{I.}~\bibnamefont{{Wasserman}}},
  \bibnamefont{and} \bibinfo{author}{\bibfnamefont{M.}~\bibnamefont{{Wyman}}},
  \bibinfo{journal}{\prd} \textbf{\bibinfo{volume}{68}},
  \bibinfo{pages}{023506} (\bibinfo{year}{2003}),
  \eprint{arXiv:hep-th/0304188}.

\bibitem[{\citenamefont{{Pshirkov} and {Tuntsov}}(2009)}]{pshirkov2009}
\bibinfo{author}{\bibfnamefont{M.~S.} \bibnamefont{{Pshirkov}}}
  \bibnamefont{and} \bibinfo{author}{\bibfnamefont{A.~V.}
  \bibnamefont{{Tuntsov}}}, \bibinfo{journal}{ArXiv e-prints}
  (\bibinfo{year}{2009}), \eprint{0911.4955}.

\bibitem[{\citenamefont{{Chernoff}}(2009)}]{chernoff2009}
\bibinfo{author}{\bibfnamefont{D.~F.} \bibnamefont{{Chernoff}}},
  \bibinfo{journal}{ArXiv e-prints}  (\bibinfo{year}{2009}),
  \eprint{0908.4077}.

\bibitem[{\citenamefont{{Vanden Berk} et~al.}(2004)\citenamefont{{Vanden Berk},
  {Wilhite}, {Kron}, {Anderson}, {Brunner}, {Hall}, {Ivezi{\'c}}, {Richards},
  {Schneider}, {York} et~al.}}]{vandenberk2004}
\bibinfo{author}{\bibfnamefont{D.~E.} \bibnamefont{{Vanden Berk}}},
  \bibinfo{author}{\bibfnamefont{B.~C.} \bibnamefont{{Wilhite}}},
  \bibinfo{author}{\bibfnamefont{R.~G.} \bibnamefont{{Kron}}},
  \bibinfo{author}{\bibfnamefont{S.~F.} \bibnamefont{{Anderson}}},
  \bibinfo{author}{\bibfnamefont{R.~J.} \bibnamefont{{Brunner}}},
  \bibinfo{author}{\bibfnamefont{P.~B.} \bibnamefont{{Hall}}},
  \bibinfo{author}{\bibfnamefont{{\v Z}.}~\bibnamefont{{Ivezi{\'c}}}},
  \bibinfo{author}{\bibfnamefont{G.~T.} \bibnamefont{{Richards}}},
  \bibinfo{author}{\bibfnamefont{D.~P.} \bibnamefont{{Schneider}}},
  \bibinfo{author}{\bibfnamefont{D.~G.} \bibnamefont{{York}}},
  \bibnamefont{et~al.}, \bibinfo{journal}{\apj} \textbf{\bibinfo{volume}{601}},
  \bibinfo{pages}{692} (\bibinfo{year}{2004}), \eprint{arXiv:astro-ph/0310336}.

\bibitem[{\citenamefont{{Weinberg}}(2008)}]{weinberg2008}
\bibinfo{author}{\bibfnamefont{S.}~\bibnamefont{{Weinberg}}},
  \emph{\bibinfo{title}{{Cosmology}}} (\bibinfo{publisher}{Oxford University
  Press}, \bibinfo{year}{2008}).

\bibitem[{\citenamefont{{Komatsu} et~al.}(2009)\citenamefont{{Komatsu},
  {Dunkley}, {Nolta}, {Bennett}, {Gold}, {Hinshaw}, {Jarosik}, {Larson},
  {Limon}, {Page} et~al.}}]{komatsu2009}
\bibinfo{author}{\bibfnamefont{E.}~\bibnamefont{{Komatsu}}},
  \bibinfo{author}{\bibfnamefont{J.}~\bibnamefont{{Dunkley}}},
  \bibinfo{author}{\bibfnamefont{M.~R.} \bibnamefont{{Nolta}}},
  \bibinfo{author}{\bibfnamefont{C.~L.} \bibnamefont{{Bennett}}},
  \bibinfo{author}{\bibfnamefont{B.}~\bibnamefont{{Gold}}},
  \bibinfo{author}{\bibfnamefont{G.}~\bibnamefont{{Hinshaw}}},
  \bibinfo{author}{\bibfnamefont{N.}~\bibnamefont{{Jarosik}}},
  \bibinfo{author}{\bibfnamefont{D.}~\bibnamefont{{Larson}}},
  \bibinfo{author}{\bibfnamefont{M.}~\bibnamefont{{Limon}}},
  \bibinfo{author}{\bibfnamefont{L.}~\bibnamefont{{Page}}},
  \bibnamefont{et~al.}, \bibinfo{journal}{\apjs}
  \textbf{\bibinfo{volume}{180}}, \bibinfo{pages}{330} (\bibinfo{year}{2009}),
  \eprint{0803.0547}.

\bibitem[{\citenamefont{{Vilenkin}}(1984)}]{vilenkin1984}
\bibinfo{author}{\bibfnamefont{A.}~\bibnamefont{{Vilenkin}}},
  \bibinfo{journal}{\apjl} \textbf{\bibinfo{volume}{282}}, \bibinfo{pages}{L51}
  (\bibinfo{year}{1984}).

\bibitem[{\citenamefont{{Hogan} and {Narayan}}(1984)}]{hogan1984}
\bibinfo{author}{\bibfnamefont{C.}~\bibnamefont{{Hogan}}} \bibnamefont{and}
  \bibinfo{author}{\bibfnamefont{R.}~\bibnamefont{{Narayan}}},
  \bibinfo{journal}{\mnras} \textbf{\bibinfo{volume}{211}},
  \bibinfo{pages}{575} (\bibinfo{year}{1984}).

\bibitem[{\citenamefont{{Sazhin} and {Khlopov}}(1989)}]{sazhin1989}
\bibinfo{author}{\bibfnamefont{M.~V.} \bibnamefont{{Sazhin}}} \bibnamefont{and}
  \bibinfo{author}{\bibfnamefont{M.~Y.} \bibnamefont{{Khlopov}}},
  \bibinfo{journal}{Soviet Astronomy} \textbf{\bibinfo{volume}{33}},
  \bibinfo{pages}{98} (\bibinfo{year}{1989}).

\bibitem[{\citenamefont{{de Vries} et~al.}(2006)\citenamefont{{de Vries},
  {Becker}, and {White}}}]{devries2006}
\bibinfo{author}{\bibfnamefont{W.~H.} \bibnamefont{{de Vries}}},
  \bibinfo{author}{\bibfnamefont{R.~H.} \bibnamefont{{Becker}}},
  \bibnamefont{and} \bibinfo{author}{\bibfnamefont{R.~L.}
  \bibnamefont{{White}}}, in \emph{\bibinfo{booktitle}{Astronomical Society of
  the Pacific Conference Series}}, edited by
  \bibinfo{editor}{\bibnamefont{{C.~M.~Gaskell, I.~M.~McHardy, B.~M.~Peterson,
  \& S.~G.~Sergeev }}} (\bibinfo{year}{2006}), vol. \bibinfo{volume}{360} of
  \emph{\bibinfo{series}{Astronomical Society of the Pacific Conference
  Series}}, pp. \bibinfo{pages}{29--+}.

\bibitem[{\citenamefont{{de Vries} et~al.}(2003)\citenamefont{{de Vries},
  {Becker}, and {White}}}]{devries2003}
\bibinfo{author}{\bibfnamefont{W.~H.} \bibnamefont{{de Vries}}},
  \bibinfo{author}{\bibfnamefont{R.~H.} \bibnamefont{{Becker}}},
  \bibnamefont{and} \bibinfo{author}{\bibfnamefont{R.~L.}
  \bibnamefont{{White}}}, \bibinfo{journal}{\aj}
  \textbf{\bibinfo{volume}{126}}, \bibinfo{pages}{1217} (\bibinfo{year}{2003}),
  \eprint{arXiv:astro-ph/0306267}.

\bibitem[{\citenamefont{{Helfand} et~al.}(2001)\citenamefont{{Helfand},
  {Stone}, {Willman}, {White}, {Becker}, {Price}, {Gregg}, and
  {McMahon}}}]{helfand2001}
\bibinfo{author}{\bibfnamefont{D.~J.} \bibnamefont{{Helfand}}},
  \bibinfo{author}{\bibfnamefont{R.~P.~S.} \bibnamefont{{Stone}}},
  \bibinfo{author}{\bibfnamefont{B.}~\bibnamefont{{Willman}}},
  \bibinfo{author}{\bibfnamefont{R.~L.} \bibnamefont{{White}}},
  \bibinfo{author}{\bibfnamefont{R.~H.} \bibnamefont{{Becker}}},
  \bibinfo{author}{\bibfnamefont{T.}~\bibnamefont{{Price}}},
  \bibinfo{author}{\bibfnamefont{M.~D.} \bibnamefont{{Gregg}}},
  \bibnamefont{and} \bibinfo{author}{\bibfnamefont{R.~G.}
  \bibnamefont{{McMahon}}}, \bibinfo{journal}{\aj}
  \textbf{\bibinfo{volume}{121}}, \bibinfo{pages}{1872} (\bibinfo{year}{2001}),
  \eprint{arXiv:astro-ph/0012442}.

\bibitem[{\citenamefont{{Trevese} et~al.}(1994)\citenamefont{{Trevese}, {Kron},
  {Majewski}, {Bershady}, and {Koo}}}]{trevese1994}
\bibinfo{author}{\bibfnamefont{D.}~\bibnamefont{{Trevese}}},
  \bibinfo{author}{\bibfnamefont{R.~G.} \bibnamefont{{Kron}}},
  \bibinfo{author}{\bibfnamefont{S.~R.} \bibnamefont{{Majewski}}},
  \bibinfo{author}{\bibfnamefont{M.~A.} \bibnamefont{{Bershady}}},
  \bibnamefont{and} \bibinfo{author}{\bibfnamefont{D.~C.} \bibnamefont{{Koo}}},
  \bibinfo{journal}{\apj} \textbf{\bibinfo{volume}{433}}, \bibinfo{pages}{494}
  (\bibinfo{year}{1994}), \eprint{arXiv:astro-ph/9407003}.

\bibitem[{\citenamefont{{Cimatti} et~al.}(1993)\citenamefont{{Cimatti},
  {Zamorani}, and {Marano}}}]{cimatti1993}
\bibinfo{author}{\bibfnamefont{A.}~\bibnamefont{{Cimatti}}},
  \bibinfo{author}{\bibfnamefont{G.}~\bibnamefont{{Zamorani}}},
  \bibnamefont{and} \bibinfo{author}{\bibfnamefont{B.}~\bibnamefont{{Marano}}},
  \bibinfo{journal}{\mnras} \textbf{\bibinfo{volume}{263}},
  \bibinfo{pages}{236} (\bibinfo{year}{1993}).

\bibitem[{\citenamefont{{Netzer} and {Sheffer}}(1983)}]{netzer1983}
\bibinfo{author}{\bibfnamefont{H.}~\bibnamefont{{Netzer}}} \bibnamefont{and}
  \bibinfo{author}{\bibfnamefont{Y.}~\bibnamefont{{Sheffer}}},
  \bibinfo{journal}{\mnras} \textbf{\bibinfo{volume}{203}},
  \bibinfo{pages}{935} (\bibinfo{year}{1983}).

\bibitem[{\citenamefont{{MacLeod} et~al.}(2008)\citenamefont{{MacLeod},
  {Ivezi{\'c}}, {de Vries}, {Sesar}, and {Becker}}}]{macleod2008}
\bibinfo{author}{\bibfnamefont{C.}~\bibnamefont{{MacLeod}}},
  \bibinfo{author}{\bibfnamefont{{\v Z}.}~\bibnamefont{{Ivezi{\'c}}}},
  \bibinfo{author}{\bibfnamefont{W.}~\bibnamefont{{de Vries}}},
  \bibinfo{author}{\bibfnamefont{B.}~\bibnamefont{{Sesar}}}, \bibnamefont{and}
  \bibinfo{author}{\bibfnamefont{A.}~\bibnamefont{{Becker}}}, in
  \emph{\bibinfo{booktitle}{American Institute of Physics Conference Series}},
  edited by \bibinfo{editor}{\bibnamefont{{C.~A.~L.~Bailer-Jones}}}
  (\bibinfo{year}{2008}), vol. \bibinfo{volume}{1082} of
  \emph{\bibinfo{series}{American Institute of Physics Conference Series}}, pp.
  \bibinfo{pages}{282--286}.

\bibitem[{\citenamefont{{Kozlowski} et~al.}(2009)\citenamefont{{Kozlowski},
  {Kochanek}, {Udalski}, {Wyrzykowski}, {Soszynski}, {Szymanski}, {Kubiak},
  {Pietrzynski}, {Szewczyk}, {Ulaczyk} et~al.}}]{kozlowski2009}
\bibinfo{author}{\bibfnamefont{S.}~\bibnamefont{{Kozlowski}}},
  \bibinfo{author}{\bibfnamefont{C.~S.} \bibnamefont{{Kochanek}}},
  \bibinfo{author}{\bibfnamefont{A.}~\bibnamefont{{Udalski}}},
  \bibinfo{author}{\bibfnamefont{L.}~\bibnamefont{{Wyrzykowski}}},
  \bibinfo{author}{\bibfnamefont{I.}~\bibnamefont{{Soszynski}}},
  \bibinfo{author}{\bibfnamefont{M.~K.} \bibnamefont{{Szymanski}}},
  \bibinfo{author}{\bibfnamefont{M.}~\bibnamefont{{Kubiak}}},
  \bibinfo{author}{\bibfnamefont{G.}~\bibnamefont{{Pietrzynski}}},
  \bibinfo{author}{\bibfnamefont{O.}~\bibnamefont{{Szewczyk}}},
  \bibinfo{author}{\bibfnamefont{K.}~\bibnamefont{{Ulaczyk}}},
  \bibnamefont{et~al.}, \bibinfo{journal}{ArXiv e-prints}
  (\bibinfo{year}{2009}), \eprint{0909.1326}.

\bibitem[{\citenamefont{{Sesar} et~al.}(2006)\citenamefont{{Sesar},
  {Svilkovi{\'c}}, {Ivezi{\'c}}, {Lupton}, {Munn}, {Finkbeiner}, {Steinhardt},
  {Siverd}, {Johnston}, {Knapp} et~al.}}]{sesar2006}
\bibinfo{author}{\bibfnamefont{B.}~\bibnamefont{{Sesar}}},
  \bibinfo{author}{\bibfnamefont{D.}~\bibnamefont{{Svilkovi{\'c}}}},
  \bibinfo{author}{\bibfnamefont{{\v Z}.}~\bibnamefont{{Ivezi{\'c}}}},
  \bibinfo{author}{\bibfnamefont{R.~H.} \bibnamefont{{Lupton}}},
  \bibinfo{author}{\bibfnamefont{J.~A.} \bibnamefont{{Munn}}},
  \bibinfo{author}{\bibfnamefont{D.}~\bibnamefont{{Finkbeiner}}},
  \bibinfo{author}{\bibfnamefont{W.}~\bibnamefont{{Steinhardt}}},
  \bibinfo{author}{\bibfnamefont{R.}~\bibnamefont{{Siverd}}},
  \bibinfo{author}{\bibfnamefont{D.~E.} \bibnamefont{{Johnston}}},
  \bibinfo{author}{\bibfnamefont{G.~R.} \bibnamefont{{Knapp}}},
  \bibnamefont{et~al.}, \bibinfo{journal}{\aj} \textbf{\bibinfo{volume}{131}},
  \bibinfo{pages}{2801} (\bibinfo{year}{2006}),
  \eprint{arXiv:astro-ph/0403319}.

\bibitem[{\citenamefont{{Giveon} et~al.}(1999)\citenamefont{{Giveon}, {Maoz},
  {Kaspi}, {Netzer}, and {Smith}}}]{giveon1999}
\bibinfo{author}{\bibfnamefont{U.}~\bibnamefont{{Giveon}}},
  \bibinfo{author}{\bibfnamefont{D.}~\bibnamefont{{Maoz}}},
  \bibinfo{author}{\bibfnamefont{S.}~\bibnamefont{{Kaspi}}},
  \bibinfo{author}{\bibfnamefont{H.}~\bibnamefont{{Netzer}}}, \bibnamefont{and}
  \bibinfo{author}{\bibfnamefont{P.~S.} \bibnamefont{{Smith}}},
  \bibinfo{journal}{\mnras} \textbf{\bibinfo{volume}{306}},
  \bibinfo{pages}{637} (\bibinfo{year}{1999}), \eprint{arXiv:astro-ph/9902254}.

\bibitem[{\citenamefont{{Uzan} and {Bernardeau}}(2001)}]{uzan2001}
\bibinfo{author}{\bibfnamefont{J.}~\bibnamefont{{Uzan}}} \bibnamefont{and}
  \bibinfo{author}{\bibfnamefont{F.}~\bibnamefont{{Bernardeau}}},
  \bibinfo{journal}{\prd} \textbf{\bibinfo{volume}{63}},
  \bibinfo{pages}{023004} (\bibinfo{year}{2001}),
  \eprint{arXiv:astro-ph/0004105}.

\bibitem[{\citenamefont{{Bernardeau} and {Uzan}}(2001)}]{bernardeau2001}
\bibinfo{author}{\bibfnamefont{F.}~\bibnamefont{{Bernardeau}}}
  \bibnamefont{and} \bibinfo{author}{\bibfnamefont{J.}~\bibnamefont{{Uzan}}},
  \bibinfo{journal}{\prd} \textbf{\bibinfo{volume}{63}},
  \bibinfo{pages}{023005} (\bibinfo{year}{2001}),
  \eprint{arXiv:astro-ph/0004102}.

\bibitem[{\citenamefont{{Kuijken} et~al.}(2008)\citenamefont{{Kuijken},
  {Siemens}, and {Vachaspati}}}]{kuijken2008}
\bibinfo{author}{\bibfnamefont{K.}~\bibnamefont{{Kuijken}}},
  \bibinfo{author}{\bibfnamefont{X.}~\bibnamefont{{Siemens}}},
  \bibnamefont{and}
  \bibinfo{author}{\bibfnamefont{T.}~\bibnamefont{{Vachaspati}}},
  \bibinfo{journal}{\mnras} \textbf{\bibinfo{volume}{384}},
  \bibinfo{pages}{161} (\bibinfo{year}{2008}), \eprint{0707.2971}.

\bibitem[{\citenamefont{{York} et~al.}(2000)\citenamefont{{York}, {Adelman},
  {Anderson}, {Anderson}, {Annis}, {Bahcall}, {Bakken}, {Barkhouser},
  {Bastian}, {Berman} et~al.}}]{york2000}
\bibinfo{author}{\bibfnamefont{D.~G.} \bibnamefont{{York}}},
  \bibinfo{author}{\bibfnamefont{J.}~\bibnamefont{{Adelman}}},
  \bibinfo{author}{\bibfnamefont{J.~E.} \bibnamefont{{Anderson}},
  \bibfnamefont{Jr.}}, \bibinfo{author}{\bibfnamefont{S.~F.}
  \bibnamefont{{Anderson}}},
  \bibinfo{author}{\bibfnamefont{J.}~\bibnamefont{{Annis}}},
  \bibinfo{author}{\bibfnamefont{N.~A.} \bibnamefont{{Bahcall}}},
  \bibinfo{author}{\bibfnamefont{J.~A.} \bibnamefont{{Bakken}}},
  \bibinfo{author}{\bibfnamefont{R.}~\bibnamefont{{Barkhouser}}},
  \bibinfo{author}{\bibfnamefont{S.}~\bibnamefont{{Bastian}}},
  \bibinfo{author}{\bibfnamefont{E.}~\bibnamefont{{Berman}}},
  \bibnamefont{et~al.}, \bibinfo{journal}{\aj} \textbf{\bibinfo{volume}{120}},
  \bibinfo{pages}{1579} (\bibinfo{year}{2000}),
  \eprint{arXiv:astro-ph/0006396}.

\bibitem[{\citenamefont{{Schneider} et~al.}(2003)\citenamefont{{Schneider},
  {Fan}, {Hall}, {Jester}, {Richards}, {Stoughton}, {Strauss}, {SubbaRao},
  {Vanden Berk}, {Anderson} et~al.}}]{schneider2003}
\bibinfo{author}{\bibfnamefont{D.~P.} \bibnamefont{{Schneider}}},
  \bibinfo{author}{\bibfnamefont{X.}~\bibnamefont{{Fan}}},
  \bibinfo{author}{\bibfnamefont{P.~B.} \bibnamefont{{Hall}}},
  \bibinfo{author}{\bibfnamefont{S.}~\bibnamefont{{Jester}}},
  \bibinfo{author}{\bibfnamefont{G.~T.} \bibnamefont{{Richards}}},
  \bibinfo{author}{\bibfnamefont{C.}~\bibnamefont{{Stoughton}}},
  \bibinfo{author}{\bibfnamefont{M.~A.} \bibnamefont{{Strauss}}},
  \bibinfo{author}{\bibfnamefont{M.}~\bibnamefont{{SubbaRao}}},
  \bibinfo{author}{\bibfnamefont{D.~E.} \bibnamefont{{Vanden Berk}}},
  \bibinfo{author}{\bibfnamefont{S.~F.} \bibnamefont{{Anderson}}},
  \bibnamefont{et~al.}, \bibinfo{journal}{\aj} \textbf{\bibinfo{volume}{126}},
  \bibinfo{pages}{2579} (\bibinfo{year}{2003}),
  \eprint{arXiv:astro-ph/0308443}.

\bibitem[{\citenamefont{{Schneider} et~al.}(2005)\citenamefont{{Schneider},
  {Hall}, {Richards}, {Vanden Berk}, {Anderson}, {Fan}, {Jester}, {Stoughton},
  {Strauss}, {SubbaRao} et~al.}}]{schneider2005}
\bibinfo{author}{\bibfnamefont{D.~P.} \bibnamefont{{Schneider}}},
  \bibinfo{author}{\bibfnamefont{P.~B.} \bibnamefont{{Hall}}},
  \bibinfo{author}{\bibfnamefont{G.~T.} \bibnamefont{{Richards}}},
  \bibinfo{author}{\bibfnamefont{D.~E.} \bibnamefont{{Vanden Berk}}},
  \bibinfo{author}{\bibfnamefont{S.~F.} \bibnamefont{{Anderson}}},
  \bibinfo{author}{\bibfnamefont{X.}~\bibnamefont{{Fan}}},
  \bibinfo{author}{\bibfnamefont{S.}~\bibnamefont{{Jester}}},
  \bibinfo{author}{\bibfnamefont{C.}~\bibnamefont{{Stoughton}}},
  \bibinfo{author}{\bibfnamefont{M.~A.} \bibnamefont{{Strauss}}},
  \bibinfo{author}{\bibfnamefont{M.}~\bibnamefont{{SubbaRao}}},
  \bibnamefont{et~al.}, \bibinfo{journal}{\aj} \textbf{\bibinfo{volume}{130}},
  \bibinfo{pages}{367} (\bibinfo{year}{2005}), \eprint{arXiv:astro-ph/0503679}.

\bibitem[{\citenamefont{{Schneider} et~al.}(2007)\citenamefont{{Schneider},
  {Hall}, {Richards}, {Strauss}, {Vanden Berk}, {Anderson}, {Brandt}, {Fan},
  {Jester}, {Gray} et~al.}}]{schneider2007}
\bibinfo{author}{\bibfnamefont{D.~P.} \bibnamefont{{Schneider}}},
  \bibinfo{author}{\bibfnamefont{P.~B.} \bibnamefont{{Hall}}},
  \bibinfo{author}{\bibfnamefont{G.~T.} \bibnamefont{{Richards}}},
  \bibinfo{author}{\bibfnamefont{M.~A.} \bibnamefont{{Strauss}}},
  \bibinfo{author}{\bibfnamefont{D.~E.} \bibnamefont{{Vanden Berk}}},
  \bibinfo{author}{\bibfnamefont{S.~F.} \bibnamefont{{Anderson}}},
  \bibinfo{author}{\bibfnamefont{W.~N.} \bibnamefont{{Brandt}}},
  \bibinfo{author}{\bibfnamefont{X.}~\bibnamefont{{Fan}}},
  \bibinfo{author}{\bibfnamefont{S.}~\bibnamefont{{Jester}}},
  \bibinfo{author}{\bibfnamefont{J.}~\bibnamefont{{Gray}}},
  \bibnamefont{et~al.}, \bibinfo{journal}{\aj} \textbf{\bibinfo{volume}{134}},
  \bibinfo{pages}{102} (\bibinfo{year}{2007}), \eprint{0704.0806}.

\bibitem[{\citenamefont{{Bentz} et~al.}(2009)\citenamefont{{Bentz}, {Peterson},
  {Netzer}, {Pogge}, and {Vestergaard}}}]{bentz2009}
\bibinfo{author}{\bibfnamefont{M.~C.} \bibnamefont{{Bentz}}},
  \bibinfo{author}{\bibfnamefont{B.~M.} \bibnamefont{{Peterson}}},
  \bibinfo{author}{\bibfnamefont{H.}~\bibnamefont{{Netzer}}},
  \bibinfo{author}{\bibfnamefont{R.~W.} \bibnamefont{{Pogge}}},
  \bibnamefont{and}
  \bibinfo{author}{\bibfnamefont{M.}~\bibnamefont{{Vestergaard}}},
  \bibinfo{journal}{\apj} \textbf{\bibinfo{volume}{697}}, \bibinfo{pages}{160}
  (\bibinfo{year}{2009}), \eprint{0812.2283}.

\bibitem[{\citenamefont{{Melnikov} and {Shevchenko}}(2008)}]{melnikov2008}
\bibinfo{author}{\bibfnamefont{A.~V.} \bibnamefont{{Melnikov}}}
  \bibnamefont{and} \bibinfo{author}{\bibfnamefont{I.~I.}
  \bibnamefont{{Shevchenko}}}, \bibinfo{journal}{\mnras}
  \textbf{\bibinfo{volume}{389}}, \bibinfo{pages}{478} (\bibinfo{year}{2008}),
  \eprint{0705.0583}.

\bibitem[{\citenamefont{{Peterson} et~al.}(2004)\citenamefont{{Peterson},
  {Ferrarese}, {Gilbert}, {Kaspi}, {Malkan}, {Maoz}, {Merritt}, {Netzer},
  {Onken}, {Pogge} et~al.}}]{peterson2004}
\bibinfo{author}{\bibfnamefont{B.~M.} \bibnamefont{{Peterson}}},
  \bibinfo{author}{\bibfnamefont{L.}~\bibnamefont{{Ferrarese}}},
  \bibinfo{author}{\bibfnamefont{K.~M.} \bibnamefont{{Gilbert}}},
  \bibinfo{author}{\bibfnamefont{S.}~\bibnamefont{{Kaspi}}},
  \bibinfo{author}{\bibfnamefont{M.~A.} \bibnamefont{{Malkan}}},
  \bibinfo{author}{\bibfnamefont{D.}~\bibnamefont{{Maoz}}},
  \bibinfo{author}{\bibfnamefont{D.}~\bibnamefont{{Merritt}}},
  \bibinfo{author}{\bibfnamefont{H.}~\bibnamefont{{Netzer}}},
  \bibinfo{author}{\bibfnamefont{C.~A.} \bibnamefont{{Onken}}},
  \bibinfo{author}{\bibfnamefont{R.~W.} \bibnamefont{{Pogge}}},
  \bibnamefont{et~al.}, \bibinfo{journal}{\apj} \textbf{\bibinfo{volume}{613}},
  \bibinfo{pages}{682} (\bibinfo{year}{2004}), \eprint{arXiv:astro-ph/0407299}.

\bibitem[{\citenamefont{{Wandel} et~al.}(1999)\citenamefont{{Wandel},
  {Peterson}, and {Malkan}}}]{wandel1999}
\bibinfo{author}{\bibfnamefont{A.}~\bibnamefont{{Wandel}}},
  \bibinfo{author}{\bibfnamefont{B.~M.} \bibnamefont{{Peterson}}},
  \bibnamefont{and} \bibinfo{author}{\bibfnamefont{M.~A.}
  \bibnamefont{{Malkan}}}, \bibinfo{journal}{\apj}
  \textbf{\bibinfo{volume}{526}}, \bibinfo{pages}{579} (\bibinfo{year}{1999}),
  \eprint{arXiv:astro-ph/9905224}.

\bibitem[{\citenamefont{{Kaspi} et~al.}(2000)\citenamefont{{Kaspi}, {Smith},
  {Netzer}, {Maoz}, {Jannuzi}, and {Giveon}}}]{kaspi2000}
\bibinfo{author}{\bibfnamefont{S.}~\bibnamefont{{Kaspi}}},
  \bibinfo{author}{\bibfnamefont{P.~S.} \bibnamefont{{Smith}}},
  \bibinfo{author}{\bibfnamefont{H.}~\bibnamefont{{Netzer}}},
  \bibinfo{author}{\bibfnamefont{D.}~\bibnamefont{{Maoz}}},
  \bibinfo{author}{\bibfnamefont{B.~T.} \bibnamefont{{Jannuzi}}},
  \bibnamefont{and} \bibinfo{author}{\bibfnamefont{U.}~\bibnamefont{{Giveon}}},
  \bibinfo{journal}{\apj} \textbf{\bibinfo{volume}{533}}, \bibinfo{pages}{631}
  (\bibinfo{year}{2000}), \eprint{arXiv:astro-ph/9911476}.

\bibitem[{\citenamefont{{Kochanek}}(2004)}]{kochanek2004}
\bibinfo{author}{\bibfnamefont{C.~S.} \bibnamefont{{Kochanek}}},
  \bibinfo{journal}{\apj} \textbf{\bibinfo{volume}{605}}, \bibinfo{pages}{58}
  (\bibinfo{year}{2004}), \eprint{arXiv:astro-ph/0307422}.

\bibitem[{\citenamefont{{Floyd} et~al.}(2008)\citenamefont{{Floyd}, {Bate}, and
  {Webster}}}]{floyd2008}
\bibinfo{author}{\bibfnamefont{D.~J.~E.} \bibnamefont{{Floyd}}},
  \bibinfo{author}{\bibfnamefont{N.~F.} \bibnamefont{{Bate}}},
  \bibnamefont{and} \bibinfo{author}{\bibfnamefont{R.~L.}
  \bibnamefont{{Webster}}}, \bibinfo{journal}{Memorie della Societa Astronomica
  Italiana} \textbf{\bibinfo{volume}{79}}, \bibinfo{pages}{1271}
  (\bibinfo{year}{2008}).

\bibitem[{\citenamefont{{Vakulik} et~al.}(2007)\citenamefont{{Vakulik},
  {Schild}, {Smirnov}, {Dudinov}, and {Tsvetkova}}}]{vakulik2007}
\bibinfo{author}{\bibfnamefont{V.~G.} \bibnamefont{{Vakulik}}},
  \bibinfo{author}{\bibfnamefont{R.~E.} \bibnamefont{{Schild}}},
  \bibinfo{author}{\bibfnamefont{G.~V.} \bibnamefont{{Smirnov}}},
  \bibinfo{author}{\bibfnamefont{V.~N.} \bibnamefont{{Dudinov}}},
  \bibnamefont{and} \bibinfo{author}{\bibfnamefont{V.~S.}
  \bibnamefont{{Tsvetkova}}}, \bibinfo{journal}{\mnras}
  \textbf{\bibinfo{volume}{382}}, \bibinfo{pages}{819} (\bibinfo{year}{2007}),
  \eprint{0708.1082}.

\bibitem[{\citenamefont{{Wayth} et~al.}(2005)\citenamefont{{Wayth}, {O'Dowd},
  and {Webster}}}]{wayth2005}
\bibinfo{author}{\bibfnamefont{R.~B.} \bibnamefont{{Wayth}}},
  \bibinfo{author}{\bibfnamefont{M.}~\bibnamefont{{O'Dowd}}}, \bibnamefont{and}
  \bibinfo{author}{\bibfnamefont{R.~L.} \bibnamefont{{Webster}}},
  \bibinfo{journal}{\mnras} \textbf{\bibinfo{volume}{359}},
  \bibinfo{pages}{561} (\bibinfo{year}{2005}), \eprint{arXiv:astro-ph/0502396}.

\bibitem[{\citenamefont{{Wyithe} et~al.}(2002)\citenamefont{{Wyithe}, {Agol},
  and {Fluke}}}]{wyithe2002}
\bibinfo{author}{\bibfnamefont{J.~S.~B.} \bibnamefont{{Wyithe}}},
  \bibinfo{author}{\bibfnamefont{E.}~\bibnamefont{{Agol}}}, \bibnamefont{and}
  \bibinfo{author}{\bibfnamefont{C.~J.} \bibnamefont{{Fluke}}},
  \bibinfo{journal}{\mnras} \textbf{\bibinfo{volume}{331}},
  \bibinfo{pages}{1041} (\bibinfo{year}{2002}),
  \eprint{arXiv:astro-ph/0112281}.

\bibitem[{\citenamefont{{Jackson} et~al.}(2005)\citenamefont{{Jackson},
  {Jones}, and {Polchinski}}}]{jackson2005}
\bibinfo{author}{\bibfnamefont{M.~G.} \bibnamefont{{Jackson}}},
  \bibinfo{author}{\bibfnamefont{N.~T.} \bibnamefont{{Jones}}},
  \bibnamefont{and}
  \bibinfo{author}{\bibfnamefont{J.}~\bibnamefont{{Polchinski}}},
  \bibinfo{journal}{Journal of High Energy Physics}
  \textbf{\bibinfo{volume}{10}}, \bibinfo{pages}{13} (\bibinfo{year}{2005}),
  \eprint{arXiv:hep-th/0405229}.

\bibitem[{\citenamefont{{Sakellariadou}}(2005)}]{sakellariadou2005}
\bibinfo{author}{\bibfnamefont{M.}~\bibnamefont{{Sakellariadou}}},
  \bibinfo{journal}{Journal of Cosmology and Astro-Particle Physics}
  \textbf{\bibinfo{volume}{4}}, \bibinfo{pages}{3} (\bibinfo{year}{2005}),
  \eprint{arXiv:hep-th/0410234}.

\bibitem[{\citenamefont{{LSST Science Collaborations: Paul A.~Abell}
  et~al.}(2009)\citenamefont{{LSST Science Collaborations: Paul A.~Abell},
  {Allison}, {Anderson}, {Andrew}, {Angel}, {Armus}, {Arnett}, {Asztalos},
  {Axelrod}, {Bailey} et~al.}}]{lsst2009}
\bibinfo{author}{\bibnamefont{{LSST Science Collaborations: Paul A.~Abell}}},
  \bibinfo{author}{\bibfnamefont{J.}~\bibnamefont{{Allison}}},
  \bibinfo{author}{\bibfnamefont{S.~F.} \bibnamefont{{Anderson}}},
  \bibinfo{author}{\bibfnamefont{J.~R.} \bibnamefont{{Andrew}}},
  \bibinfo{author}{\bibfnamefont{J.~R.~P.} \bibnamefont{{Angel}}},
  \bibinfo{author}{\bibfnamefont{L.}~\bibnamefont{{Armus}}},
  \bibinfo{author}{\bibfnamefont{D.}~\bibnamefont{{Arnett}}},
  \bibinfo{author}{\bibfnamefont{S.~J.} \bibnamefont{{Asztalos}}},
  \bibinfo{author}{\bibfnamefont{T.~S.} \bibnamefont{{Axelrod}}},
  \bibinfo{author}{\bibfnamefont{S.}~\bibnamefont{{Bailey}}},
  \bibnamefont{et~al.}, \bibinfo{journal}{ArXiv e-prints}
  (\bibinfo{year}{2009}), \eprint{0912.0201}.

\bibitem[{\citenamefont{{Perryman} et~al.}(2001)\citenamefont{{Perryman}, {de
  Boer}, {Gilmore}, {H{\o}g}, {Lattanzi}, {Lindegren}, {Luri}, {Mignard},
  {Pace}, and {de Zeeuw}}}]{perryman2001}
\bibinfo{author}{\bibfnamefont{M.~A.~C.} \bibnamefont{{Perryman}}},
  \bibinfo{author}{\bibfnamefont{K.~S.} \bibnamefont{{de Boer}}},
  \bibinfo{author}{\bibfnamefont{G.}~\bibnamefont{{Gilmore}}},
  \bibinfo{author}{\bibfnamefont{E.}~\bibnamefont{{H{\o}g}}},
  \bibinfo{author}{\bibfnamefont{M.~G.} \bibnamefont{{Lattanzi}}},
  \bibinfo{author}{\bibfnamefont{L.}~\bibnamefont{{Lindegren}}},
  \bibinfo{author}{\bibfnamefont{X.}~\bibnamefont{{Luri}}},
  \bibinfo{author}{\bibfnamefont{F.}~\bibnamefont{{Mignard}}},
  \bibinfo{author}{\bibfnamefont{O.}~\bibnamefont{{Pace}}}, \bibnamefont{and}
  \bibinfo{author}{\bibfnamefont{P.~T.} \bibnamefont{{de Zeeuw}}},
  \bibinfo{journal}{\aap} \textbf{\bibinfo{volume}{369}}, \bibinfo{pages}{339}
  (\bibinfo{year}{2001}), \eprint{arXiv:astro-ph/0101235}.

\end{thebibliography}
\label{lastpage}

\end{document}